\documentclass[preprint,showpacs,preprintnumbers,amsmath,amssymb]{revtex4}

\usepackage{graphicx}
\usepackage{dcolumn}
\usepackage{bm}

\newcommand{\lw}[1]{\smash{\lower 1.5ex\hbox{#1}}}
\newcommand{\mapright}[1]
{\smash{\mathop{\hbox to 1cm{\rightarrowfill}}\limits^{#1}}}

\begin{document}

\title{Analysis of distribution of cosmic microwave background photon in terms 
of non-extensive statistics and formulas with temperature fluctuation}

\author{M. Biyajima}
\author{M. Kaneyama}
\author{Y. Kurashima}
\author{T. Yamashita}
\affiliation{
Department of physics, Shinshu University, Matsumoto 390-0821, Japan
}

\author{T. Mizoguchi}
\affiliation{
Toba National College of Maritime Technology, Toba 517-8501, Japan
}

\date{\today}

\begin{abstract}
To take into account the temperature fluctuation in the Planck distribution, we 
calculate convolution integral with several probability distributions. Using 
these formula as well the Planck distribution and a formula in the 
non-extensive statistics, we analyze the data measured by the Cosmic Background 
Explorer (COBE). Our analysis reveals that the derivation from the Planck 
distribution is estimated as $|q-1| = 4.4\times10^{-5}$, where $q$ means the 
magnitude of the non-extensivity or the temperature fluctuation, provided that 
the dimensionless chemical potential proposed by Zeldovich and Sunyaev exists. 
Comparisons of new formulas and the Planck distribution including the 
Sunyaev-Zeldovich (S-Z) effect are made.
\end{abstract}


\maketitle

\section{\label{sec:1}Introduction}
One of interesting subjects in thermodynamics is relating to the non-extensive 
statistics~\cite{tsallis1,tsallis2,tsallis3}. Several years ago, authors of 
Refs.~\cite{wilk,beck} showed that formulas in the non-extensive statistics, 
i.e., the temperature fluctuations, are calculated by the convolution integral 
with the gamma distribution~\cite{wilk} and a calculation with a distribution 
described by $\exp(-2|u|^{\alpha})$, where $\alpha$ is fractional 
number~\cite{beck}. In this report, we calculate the temperature fluctuation 
in the Planck distribution
\begin{equation}
  D_{Planck}(\beta,\nu,\mu)=\frac{C_B \nu^3}{e^{\beta \omega+\mu}-1}
  \label{d-planck}\:,
\end{equation}
where $C_B = 8\pi h/c^3$, $\omega=h\nu$ and $\beta =1/k_BT$. $\mu$ denotes the 
dimensionless chemical potential introduced by Zeldovich and 
Sunyaev~\cite{zeldovich}. Equation~\eqref{d-planck} is applied to the data 
measured by the Cosmic Background Explorer~(COBE)~\cite{mather1,mather2}. 

The formula in the non-extensive statistics is obtained in Ref.~\cite{tsallis2} 
as
\begin{eqnarray}
  F^{(NETD)}(\beta,\nu)=D_{Planck}(\beta,\nu)[1-e^{-x}]^{(q-1)}\nonumber\\
  \times \Bigl\{ 1+(1-q)x\Bigl[\frac{1+e^{-x}}{1-e^{-x}}
  -\frac{x}{2}\frac{1+3e^{-x}}{(1-e^{-x})^2} \Big] \Bigr\}
  \label{netd}\:,
\end{eqnarray}
where $x=\beta\omega$.
(It should be noticed that calculation in Ref.~\cite{tsallis2} is done without 
$\mu$. The value of $\chi^2$ is 108 and the degree of freedom is 
31~\cite{tsallis2}).

In the next section, we calculate the convolution integrals using the several 
probability distributions~\cite{biyajima,glauber,goel}. In the 3rd section we 
analyze the data reported in Ref.~\cite{fixsen,mather3} in terms of 
Eqs.~\eqref{d-planck} and~\eqref{netd} with $\mu$, as well as new formulas. In 
the final section, concluding remarks are presented. We look for the origin of 
the temperature fluctuation.\\

\section{\label{sec:2}Convolution integrals of Eq.~(\ref{d-planck}) with 
probability distributions}
First of all, according to Refs.~\cite{wilk,beck}, we consider the gamma 
distribution for the temperature fluctuation as
\begin{equation}
  P(\alpha,\beta,\beta_0)=\frac{1}{\Gamma(\alpha)}
  \Bigl(\frac{\alpha}{\beta_0}\Bigr)^{\alpha}
  \beta^{\alpha-1}e^{-\frac{\alpha}{\beta_0}\beta}
  \label{gamma}
\end{equation}
The formula, Eqs.~\eqref{gamma}, is also named the $\chi^2$ distribution. The 
convolution integral is given as\\
\begin{widetext}
\begin{eqnarray}
  F^{(gamma)}(\beta_{0},\nu,\alpha)=\int^{\infty}_{0}d\beta 
  P(\alpha,\beta,\beta_{0})\times D_{Planck}(\beta,\nu,\mu)
  =\sum^{\infty}_{n=1}C_{B}\nu^3 e^{-n\mu} \frac{1}{(1+\frac{\beta_0}
  {\alpha}\omega n)^{\alpha}}\nonumber\\
  \mapright{\alpha \gg 1}\sum^{\infty}_{n=1}C_{B}\nu^{3}
  e^{-n\mu-\beta_0\omega n+\frac{1}{2}\beta_0^2 \omega^2n^2(q-1)+O((q-1)^2)}
  \label{gamma2}\:,
\end{eqnarray}
\end{widetext}
where $\alpha =\frac{1}{q-1}$ and $(q-1) \ll 1$ are assumed.

The following generalized $\chi^2$ distribution is known as the non-central 
$\chi^2$ distribution or the generalized Glauker-Lachs formula and 
Perina-McGill formula~\cite{biyajima,glauber,goel}.
\begin{eqnarray}
  \Bigr(\frac{\alpha}{p}\Bigl)\Biggr[ 
  \frac{\Bigl(\frac{\beta}{\beta_0}\Bigl)}{(1-p)} \Biggl]^{\frac{\alpha-1}{2}}
  \exp \Bigl[ -\frac{\alpha (1-p)}{p}-\frac{\alpha}{p}\frac{\beta}{\beta_0} 
  \Bigr]\nonumber\\
  \times I_{(\alpha-1)}\Biggr( 2\sqrt{\frac{\beta}{\beta_{0}}
  \Bigr( \frac{\alpha}{p} \Bigr)^{2}(1-p)}\Biggr)
\label{chi}\:,
\end{eqnarray}
where $p=\langle n_{th}\rangle/\langle n_{total}\rangle$ and $I$ denotes the 
modified Bessel function. $\langle n_{th}\rangle$ does the thermal boson. 
Eq.~\eqref{chi} is the solution of the following Fokken-Planck equation
\begin{equation}
  \frac{\partial P}{\partial t}=
  -\frac{\partial}{\partial \beta}[-\frac{1}{2}\alpha \eta (\beta-1)] P
  +\frac{1}{2}\frac{\partial^2}{\partial \beta^2}\eta \beta P
  \label{F-P}\:,
\end{equation}
where $\alpha$ and $\eta$ are parameters. This is named the stochastic process 
for birth and death process with the immigration. The convolution integral of 
Eqs.~\eqref{d-planck} and \eqref{chi} is expressed as
\begin{eqnarray}
  &&F^{(NCC)}(\beta_{0},\nu,\alpha,\mu,p)\nonumber\\
  &&=\sum^{\infty}_{n=0}C_{B}\nu^{3}e^{-n\mu}
  \frac{1}{(1+\frac{\beta_{0}}{\alpha}\omega n)^{\alpha}}
  e^{-\frac{\alpha(1-p)}{p}+\kappa}
  \label{ncc}\:,
\end{eqnarray}
where 
$$
\kappa =\Bigr[ \beta_{0}\Bigl(\frac{\alpha}{p}\Bigr)^{2}(1-p) \Bigl]
\Biggr{/}\Bigl[ \frac{\alpha}{p\beta_0}+\omega n \Bigr].
$$
As $p\to 1$ and $\alpha =\frac{1}{q-1}\gg 1$, Eq.~\eqref{ncc} reduces to 
Eq.~\eqref{gamma2}.

As the third case, we consider the Gaussian distribution.
\begin{equation}
  P(V,\beta,\beta_0)=\frac{1}{\sqrt{2 \pi V^2}}
  \exp \Bigl[ -\frac{(\beta - \beta_0)^2}{2V^2}\Bigr]
  \label{gauss}\:,
\end{equation}
where $V^2$ denotes the variance. This distribution is relating to the solution 
of the Ornstein-Uhlenbeck stochastic process,
\begin{equation}
  \frac{\partial P}{\partial t}=
  \frac{\partial}{\partial \beta}r\beta P+\frac{\sigma^2}{2}
  \frac{\partial^{2} P}{\partial \beta^2}
  \label{O-U}\:.
\end{equation}
From the convolution integral of Eqs.~\eqref{d-planck} and~\eqref{gauss}, we 
obtain the following formula
\begin{equation}
  F^{(Gauss)}(\beta_{0},\nu,\alpha,\mu,V)=\sum^{\infty}_{n=1}C_{B}\nu^{3}
  e^{-n\mu}e^{-n\beta_{0}\omega +\frac{n^2 \omega^2}{2}V^2}
  \label{f-gauss}\:.
\end{equation}
When $V^2$ is expressed by $\beta_0^2 (q-1)$, 
i.e., the temperature fluctuation, Eq.~\eqref{f-gauss} reduces to 
Eq.~\eqref{gamma2}

\section{\label{sec:3}Analysis of data by COBE by means of 
Eqs.~(\ref{d-planck}), (\ref{netd}), (\ref{gamma2}), (\ref{ncc}) and 
(\ref{f-gauss})}
\begin{figure}
  \includegraphics{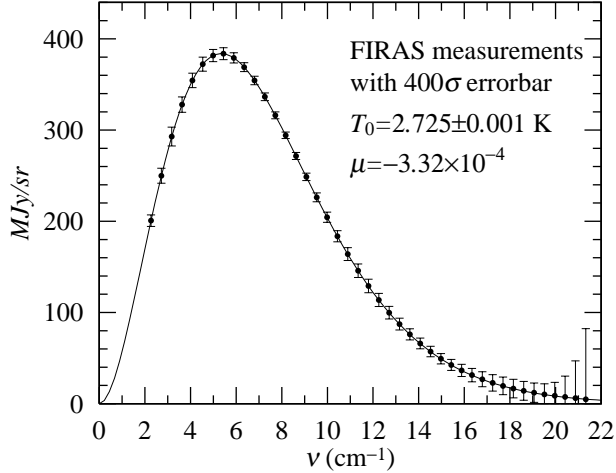}
  \caption{\label{fig:1} Analysis of data by the COBE in terms of 
  Eqs.~\eqref{d-planck}. The data sets are obtained from T.~C.~Mather and 
  D.~J.~Fixsen. See Refs~\cite{fixsen,mather3}. We use a method by 
  D.~Seibert~\cite{seibert} in calculation of error bar. The diagonal 
  components are shown.}
\end{figure}

Using the CERN-MINUIT program, we can estimate values of 
parameters contained in formulas given in previous sections. In 
Fig.~\ref{fig:1}, we have confirmed the result reported in 
Refs.~\cite{fixsen,mather3}. Our results are shown in Table~\ref{tab:1}. As 
seen in Table~\ref{tab:1}, values of $\chi^2$'s for Eqs.~\eqref{netd} and 
\eqref{gamma2} become to be somewhat better than that of the Planck 
distribution (Eq.~\eqref{d-planck}), because of the parameter q. The magnitude 
of $(q-1)$ denotes the degree of the deviation from the Planck distribution 
including $\mu$, i.e. Eq.~\eqref{d-planck}.
\begin{table*}
  \caption{\label{tab:1}Analysis of data by the COBE by means of 
  Eqs.~\eqref{d-planck}, \eqref{netd}, \eqref{gamma2}, \eqref{ncc} and 
  \eqref{f-gauss}.}
  \begin{ruledtabular}
  \begin{tabular}{cccccc}
  formula & $T_0=1/\beta_0$ & $q-1$ & $\chi^2$/n.d.f & $\mu$ & $p$\\
  \hline
  Eq.~\eqref{d-planck}
  & $2.726\pm 0.001$ & --- & $141/41$ & $0$ (fixed) & ---\\
  (Planck)
  & $2.725\pm 0.001$ & --- & $82/40$  & $(-3.32\pm 0.43)\times 10^{-4}$ & ---\\
  \hline
  Eq.~\eqref{netd}
  & $2.726\pm 0.001$ & $(2.00\pm 0.07)\times 10^{-5}$ & $167/40$ 
  & $0$ (fixed) & ---\\
  (NETD)
  & $2.725\pm 0.001$ & $(4.67\pm 1.60)\times 10^{-5}$ & $74/39$  
  & $(-5.16\pm 0.77)\times 10^{-4}$ & ---\\
  \hline
  Eq.~\eqref{gamma2}
  & $2.726\pm 0.001$ & $(2.00\pm 0.06)\times 10^{-5}$ & $169/40$ 
  & $0$ (fixed) & ---\\
  (gamma)
  & $2.725\pm 0.001$ & $(4.40\pm 1.50)\times 10^{-5}$ & $74/39$  
  & $(-5.16\pm 0.76)\times 10^{-4}$ & ---\\
  \hline
  Eq.~\eqref{ncc}
  & $2.726\pm 0.001$ & $(-3.99\pm 0.86)\times 10^{-5}$ & $119/39$ 
  & $0$ (fixed) & $0.9995\pm 0.6842$\\
  (NCC)
  & $2.725\pm 0.001$ & $(4.40\pm 1.62)\times 10^{-5}$ & $74/38$  
  & $(-5.16\pm 0.74)\times 10^{-4}$ & $0.9995\pm 0.6641$\\
  \hline
  Eq.~\eqref{f-gauss}
  & $2.726\pm 0.001$ & $(2.00\pm 0.06)\times 10^{-5}$ & $169/40$ 
  & $0$ (fixed) & ---\\
  (Gauss)
  & $2.725\pm 0.001$ & $(4.40\pm 1.47)\times 10^{-5}$ & $74/39$  
  & $(-5.16\pm 0.74)\times 10^{-4}$ & ---\\
  \end{tabular}
  \end{ruledtabular}
\end{table*}

\section{\label{sec:4}Concluding remarks}
We have calculated Eqs.~\eqref{gamma2}, ~\eqref{ncc} and~\eqref{f-gauss} 
and applied them to the data measured by the COBE~\cite{fixsen,mather3}. The 
following remarks are obtained from our analysis.
\begin{enumerate}
  \item It is shown that the dimensionless chemical potential $\mu$ is playing 
  an important role in Eq.~\eqref{d-planck}, because of values of $\chi^2$'s.
  \item The deviations from the Planck distribution including the chemical 
  potential for the cosmic microwave background (CMB) photon is estimated by 
  $|q-1|= 4.4\times 10^{-5}$. 
  
  Both approaches, the non-extensive statistics and the convolution integrals, 
  show the same magnitude.
  \item As $(q-1) \ll 1$ and $V^2 = \beta_0 (q-1)$ are assumed, it is difficult 
  to distinguish the gamma distribution and the Gaussian distribution for the 
  temperature fluctuation.
  \item From $p\approx 1$ (in spite of large error bars) in Eq.~\eqref{ncc} in 
  Table~\ref{tab:1}, it is known that the CMB photon is the pure thermal photon.
  \begin{table*}
    \caption{\label{tab:2}Analysis of data by the COBE in terms of 
    Eq.~\eqref{S-Z}.}
    \begin{ruledtabular}
    \begin{tabular}{ccccc}
    formula & $T_0=1/\beta_0$ & $y$ & $\chi^2$/n.d.f & $\mu$\\
    \hline
    \lw{Eq.~\eqref{S-Z}}
    & $2.726\pm 0.001$ & $(-1.99\pm 0.43)\times 10^{-5}$ & $119/41$ 
    & $0$ (fixed)\\
    & $2.725\pm 0.001$ & $(2.20\pm 0.75)\times 10^{-5}$ & $74/40$ 
    & $(-5.16\pm 0.76)\times 10^{-4}$\\
    \end{tabular}
    \end{ruledtabular}
  \end{table*}
  \item To look for the origin of the magnitude of the non-extensivity or the 
  temperature fluctuation, we analyze the data by the following formula 
  including the Sunyaev-Zeldovich (S-Z) effect~\cite{zeldovich,rybicki}, 
  \begin{widetext}
  \begin{eqnarray}
    F^{(S-Z\: effect)}(x,\beta,\mu,y)
    = D_{Planck}(\beta,\nu,\mu) + C_B \nu^3\frac{xye^x}{(e^x-1)^2}
    \Biggr[ x\coth \frac x2 -4 \Biggl]
    \label{S-Z}\:,
  \end{eqnarray}
  \end{widetext}
  where $x = \beta_0\omega + \mu$. $y$ is named the $y$ parameter in the 
  Compton scattering,
  \begin{eqnarray}
    y = \int dl n_e \sigma_T \frac{k_BT_e}{m_ec^2}
    \label{y-param}\:,
  \end{eqnarray}
  where $l$, $n_e$, $\sigma_T$ and $T_e$ are the size of region with high 
  temperature in the cosmos, the number density of electrons, the cross section 
  of Thomson scattering and the temperature of electron, respectively. Our 
  estimated values are shown in Table~\ref{tab:2}. The magnitude of $y$ 
  parameter $y \approx (2.2\pm 0.75)\times 10^{-5}$ is almost the same as that 
  of $|q-1|$ in Table~\ref{tab:1}. This coincidence suggests that 
  Eqs.~\eqref{gamma2},~\eqref{ncc} and~\eqref{f-gauss} work well. 
\end{enumerate}
\begin{figure}
  \includegraphics{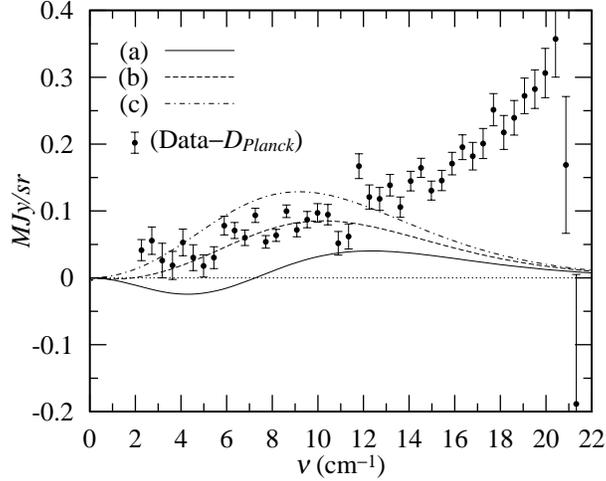}
  \caption{\label{fig:2} Correction term to the Planck distribution. Planck 
  distributions are subtracted at data points. (a) The second term of 
  Eq.~\eqref{S-Z}, i.e., Sunyaev-Zeldovich (S-Z) effect. (b) Difference between 
  Eq.~\eqref{netd} and the Planck distribution with $\mu$. (c) The same as (b) 
  but Eq.~\eqref{gamma2} and the Planck distribution with $\mu$.}
\end{figure}

We summarize the following concerning the item 5): The deviations from the 
Planck distribution for the CMB photon are seen through the non-extensivity or 
the temperature fluctuation. Moreover, the physical origin is attributed to the 
Compton scattering between electrons in the galaxies and CMB 
photons~\cite{zeldovich,rybicki}. The behavior of the second term of 
Eq.~\eqref{S-Z} is shown in Fig.~\ref{fig:2}. The crossing point on the 
horizontal line(frequency) is determined by $x\coth \frac x2 = 4$. Differences 
between Eqs.~\eqref{netd} and \eqref{gamma2} and the Planck distribution with 
$\mu$ are also shown in Fig.~\ref{fig:2}. The small quantity, (0.1-0.05)/400, 
is reflecting the physical meaning of the inverse Compton scattering in the 
galaxies. It is observed through Eqs.~\eqref{netd} and \eqref{gamma2}.

\begin{acknowledgments}
Authors would like to thank Prof.~J.~C.~Mather and Prof.~D.~J.~Fixsen for their 
kindness concerning the data sets by the COBE. Moreover, they are owing to 
Prof.~N.~Sugiyama for useful information. One of authors (M.~K.) is partially 
supported by Faculty of Science.
\end{acknowledgments}

\end{document}